\documentclass{elsart}
\usepackage{natbib}
\input{epsfig.sty}
\begin{document}
\runauthor{Gai and Schoessow}
\begin{frontmatter}
\title{Design and Simulation of a High Frequency High
Power Rf Extraction Device Using a Dielectric-Loaded
Waveguide\thanksref{X}}
\author{Wei Gai\thanksref{wgemail},}
\author{Paul Schoessow}
\thanks[X]{Work supported by US Department of Energy, High Energy Physics
Division under contract No. W-31-109-ENG-38.}
\thanks[wgemail]{email:wg@hep.anl.gov}
\address{High Energy Physics Division\\
Argonne National Laboratory\\
9700 S Cass Ave\\
Argonne IL 60439
}
\begin{abstract}
We consider the use of a dielectric-loaded structure to extract rf energy from a
high current electron drive beam as a power source for a high energy two-beam
accelerator. This represents an alternative technique which we show to have some
significant advantages over the use of the currently proposed corrugated 
metal structures
as power extraction devices. 
We discuss a particular design that will extract high power rf
(0.1 -- 1 GW) from a high current drive beam.  Rf generation and transport in
this class of
devices have already been demonstrated at lower frequencies. 
We discuss the design
parameters for 15 and 30 GHz dielectric transfer structures and some possible
experiments.

\end{abstract}
\begin{keyword}
wakefield, linear collider, rf structures
\PACS 41.75.Lx,41.60.Bq,84.40.-x,29.17.+w
\end{keyword}
\end{frontmatter}

\section{Introduction}
One trend in proposed future high energy linear collider designs is the use of 
high
accelerating gradients in room temperature high frequency structures. This approach has
the advantage that the average rf power for a given accelerating gradient scales
 as
1/(frequency)$^2$ \cite{R1}, while the breakdown field is empirically found to 
scale approximately
as (frequency)$^{1/3}$ \cite{R2}. At the same time, the technology of 
conventional (klystron) rf
sources is difficult to implement at frequencies higher than X-band (11.4 GHz) 
\cite{R1}.  Two-
beam acceleration 
\cite{R3} overcomes this difficulty by using a conventionally accelerated
high current drive beam as the rf source for the high energy accelerator.

\begin{figure*}
\epsfig{file=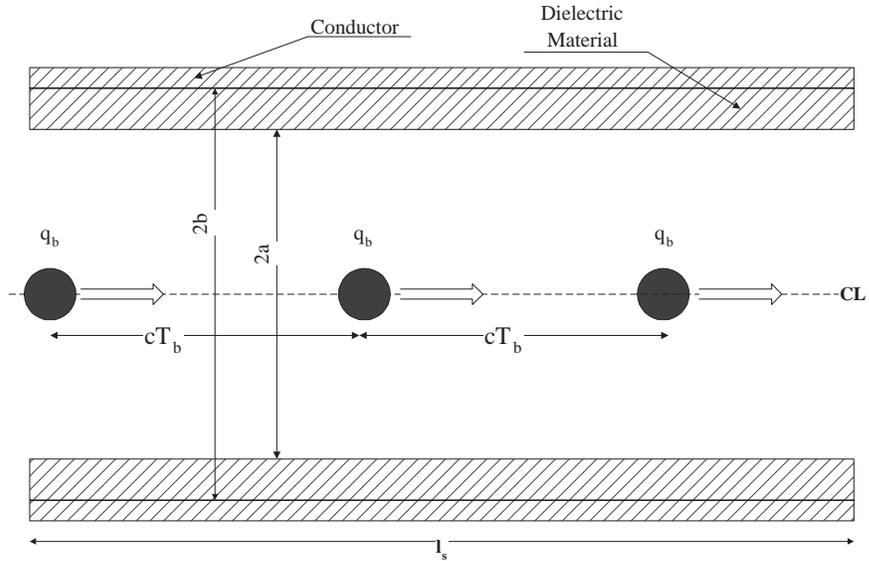,height=3in}
\caption{Dielectric loaded structure geometry. $q_b$ is the charge/bunch and
$cT_b$ is the spacing of the bunches in the train. $l_s$ is the total length
of the structure.}
\label{struct}
\end{figure*}

Two beam acceleration requires generation and propagation of a high current drive beam,
and an efficient transfer structure to extract energy from the drive beam. We propose that
a dielectric loaded waveguide represents an optimal power extraction/transfer structure.
Dielectric structures have been investigated experimentally at some length
\cite{R4,R5}. We
summarize some of the important features of this class of devices compared to
conventional copper rf structures:
\begin{itemize}
\item Simplicity of construction. As shown in figure \ref{struct}, the structure is a simple ceramic
tube which is inserted into a conducting copper sleeve.
\item Higher order mode damping. Parasitic HEM modes generated by off-axis
injection errors can be easily suppressed 
\cite{R6} by interrupting the outer conductor
with axial cuts and surrounding the sleeve with an absorbing material. Note that
this technique requires no modification to the dielectric tube and only modest
machining of the exterior conducting sleeve.
\item Recent advances in dielectric materials for 
microwave applications. Very low loss
($tan ~\delta ~\simeq~ 10^{-4}$) and high Q materials are commercially available 
\cite{R7}.
\end{itemize}

Other issues which may represent possible limitations on the performance of dielectric
structures are actively being investigated:
\begin{itemize}
\item Efficient rf coupling. 
The wakefield generated by the drive beam must be coupled
out to a rectangular waveguide for transfer to the accelerating structure. This
involves a mode conversion from TM$_{01}$ in the dielectric tube to TE in the
waveguide. Significant progress on efficient coupling design has been made 
\cite{R8}.
\item Charging. Beam halo or missteered beam impinging on the dielectric tube can
cause large electrostatic charge buildup, resulting in deflection of the beam or
volume breakdown of the structure. We have observed both these effects in our
experiments using polymer based materials such as rexolite or nylon but have not
observed this in any of the glass or ceramic materials we have tested. Changes in
microwave dielectric properties with time due to radiation damage have also not
been seen in our experiments.
\item Rf breakdown. The work in this area has been almost entirely confined to 
copper structures \cite{R2}. We have currently generated rf surface fields in 
ceramic structures
$>$ 10 MV/m at 15 GHz \cite{R5} (limited by the available drive beam properties); no
breakdown was observed at this level. We expect that significantly higher fields
can be developed at these frequencies before breakdown occurs but it is worth
pointing out that these devices are interesting as power extraction 
structures even
at 10 MeV/m scale gradients.
\end{itemize}

\section{Rf Power Generation in a Dielectric Structure}

We follow the approach and notation of reference 
\cite{R9}. Consider a dielectric structure
(figure \ref{struct}) with inner radius a and outer radius b. 
An electron bunch train passing through
the device will generate a steady state RF power

\begin{equation}\label{fo:eqn1}
P={{l_s^2 \omega}\over {4c}}({q_b \over T_b})^2 ({R' \over Q})({1\over\beta_g}-1)F^2(\sigma_z)
\end{equation}

where $l_s$ is the length of the structure, 
$q_b$ is the charge/bunch, $cT_b$ is the bunch separation,
$\beta_g$ is the group velocity and $F(\sigma_z)$ is a form factor proportional 
to the bunch spectrum
(assumed Gaussian). $R'/Q$ 
is the normalized ratio of the shunt impedance of the structure
to its quality factor and encapsulates all 
the details of the structure geometry.

To compute  $R'/Q$ for a dielectric structure, we use

\begin{equation}\label{fo:eqn2}
R'/Q = {{4 E_{z0}}\over{q_b \omega}}
\end{equation}

where $E_{z0}$ is the longitudinal decelerating wakefield in the device. 
In the limit of an
infinitely long structure, the analytic expression for the wakefield is given in
\cite{R10}. This provides $R'/Q$ via equation \ref{fo:eqn2}, 
and we can proceed to compute and optimize the properties of our structures.

Since the formalism of reference 
\cite{R10} does not allow for the effects of structure
boundaries, rf coupling etc., it only yields an approximate evaluation of the properties of
these structures.  We will also give more accurate results of numerical calculations which
include finite structure length and rf extraction effects.

\section{Design of the Dielectric Power Extraction/Transfer Structures}

In this section we give a reference design for both 15 and 30 GHz structures.
While the choice of device parameters are not identical to the
CLIC final design, this work
allows the study of the physics and engineering aspects of dielectric-
based power extraction devices. We have attempted to develop power levels
of 100 MW with large coupling to the beam
and using a common, commercially available dielectric.

While 30
GHz is the design frequency of CLIC \cite{R9},  
there are a number of advantages in
developing a 15 GHz structure first. 
The lower frequency is compatible with both the
microwave measurement equipment and the beam spectrum 
available at the AWA \cite{R5}.
Also, there is some question whether the use of 30 GHz rf does in fact 
provide significant
performance improvement over 10 GHz range devices \cite{R11}.

The choice of dielectric is Cordierite which 
has a relatively low permittivity (4.5) and
low losses (Q=5000, $tan ~\delta ~=~ 0.0002$).  This material is commercially 
available \cite{R7}, and we have verified  its dielectric properties in bench 
tests.  

The choice of dielectric is not
unique; for example, steatite, another commercially available material, has a 
permittivity
of 5.8 and loss tangent of 0.0001 and would perform equally well. Theoretically
one could also use a low coupling strength $R'/Q$ using a thin layer
of high permittivity material to line the structure \cite{wg}.

The inner radius a must be chosen to transport all the drive beam. The choice of
 dielectric
material and the device frequency then fix the outer radius b. Using the methods
 outlined
previously, we can calculate $R'/Q$ and power generated in these devices 
assuming CLIC
test facility beam parameters.

Because of the simple geometry, the structures will be easy to machine. 
Conservatively
assuming that the outer and inner radii of the structures can be machined to a 
tolerance of
25 $\mu$m, the worst case deviations from nominal frequency due to this 
tolerance are 3.4\%
for the 30 GHz device and 1.5\% for the 15 GHz structure.

\section{Numerical Calculations}

We also used the finite difference time domain program ARRAKIS 
\cite{R12} to calculate
 the
power generated in these structures while taking into account the finite 
structure length,
group velocity of the wakefield, etc. In order to simulate the effects of power
extraction
from the structure, an absorbing boundary was placed at the downstream end of 
the
dielectric tube in the computational geometry.

\begin{figure*}
\epsfig{file=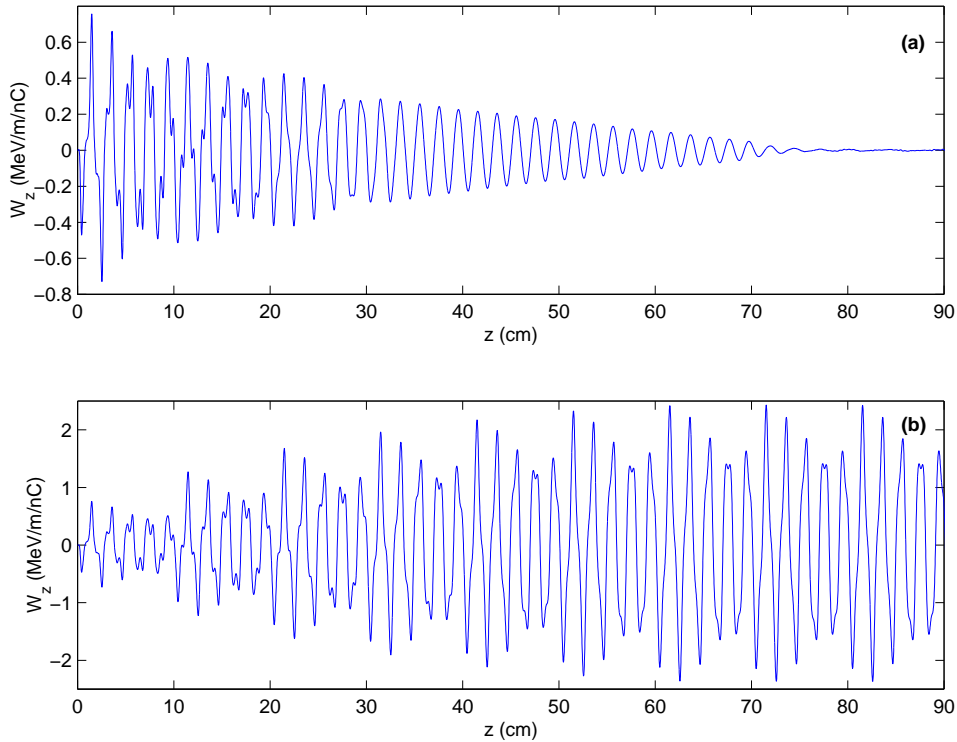,width=5in}
\caption{Numerical calculation of the longitudinal wake in the 15 GHz
dielectric structure of Table
\ref{tab1}. (a) Single bunch. (b) Bunch train. 
The bunch separation is approximately 10 cm (333 ps, 5 $\lambda_{rf}$) and
the beam intensity is normalized to 1 nC/bunch.
 }
\label{fig2}
\end{figure*}

\begin{figure*}
\epsfig{file=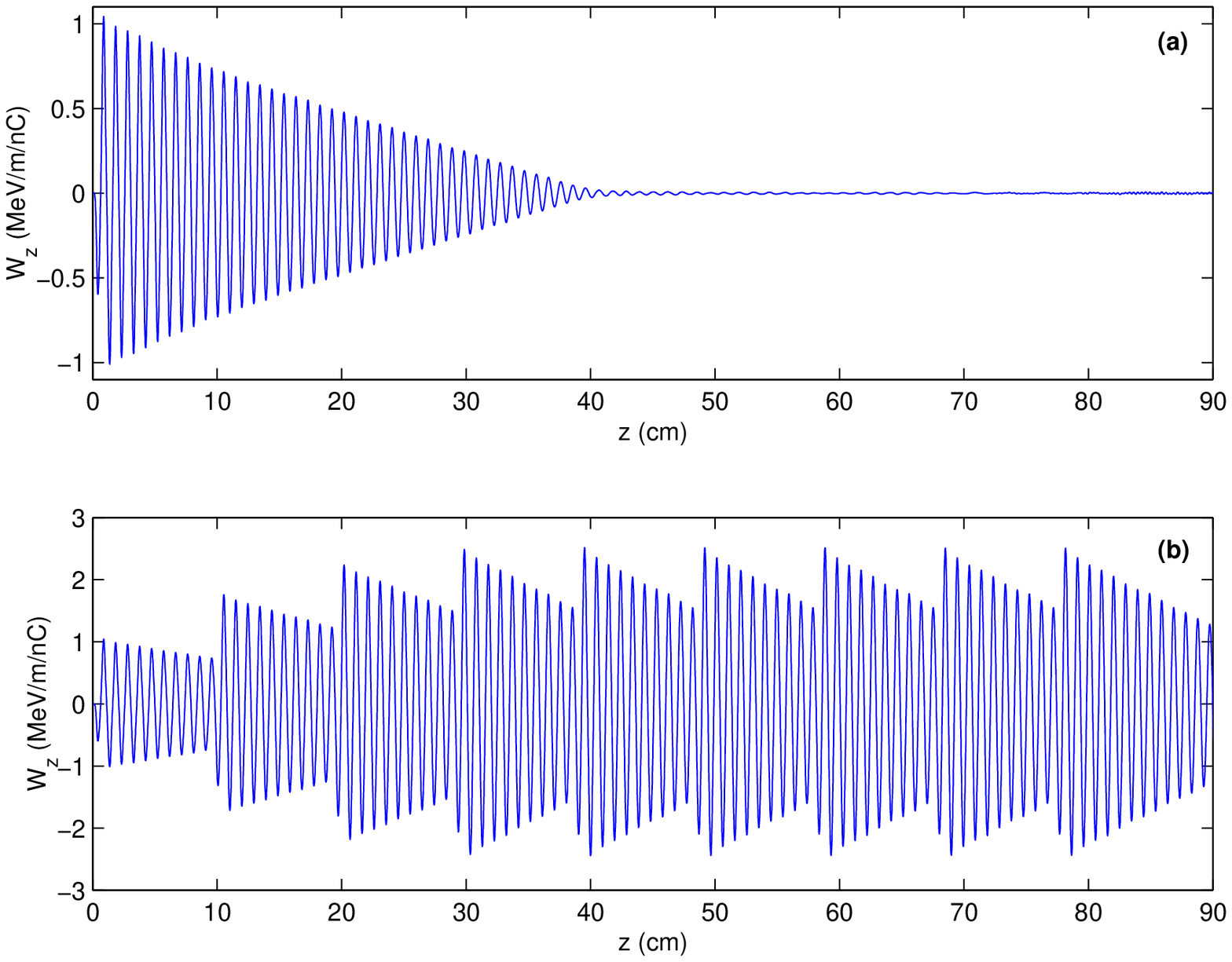,width=5in}
\caption{Numerical calculation of the longitudinal wake in the 30 GHz
dielectric structure (Table
\ref{tab1}). (a) Single bunch. (b) Bunch train. 
The bunch separation is approximately 10 cm (10 $\lambda_{rf}$).
The beam intensity is normalized to 1 nC/bunch. }
\label{fig3}
\end{figure*}

Figures \ref{fig2}a and \ref{fig3}a show the longitudinal wake potentials for the 15 and 30 GHz devices for
a single drive bunch and for a train of 1 nC bunches spaced by 10 cm. Buildup of
 fields to
equilibrium occurs relatively quickly-- after four bunches (1.33 ns) for the 15
 GHz
device and three bunches (1.0 ns) for the 30 GHz structure due to its larger 
group
velocity. The apparent ``damping" of the wakefield after each bunch is in fact 
due to the large group velocity of the fields and the finite length of the 
devices. The 15
GHz wake
shows some additional structure due to the excitation of the TM$_{02}$
mode at 42.8 GHz and
the TM$_{03}$ at 73.4 GHz. The higher frequency modes result in some 
inefficiency but in
general will not be transported out of the structure when the coupling is 
optimized for the
fundamental frequency.

The bunch train wakes shown in figures \ref{fig2}b and 
\ref{fig3}b reflect finite structure length
 and
group velocity effects by their not purely sinusoidal character. In order to 
compute the
useful power generated at the fundamental, we filter the data by performing an 
FFT and
zeroing all Fourier components outside a fixed window in $\Delta f/f_0$. 
The inverse FFT then
yields the amplitude W of the wake within that bandwidth, 
and the power generated is
\begin{equation}\label{fo:eqn3}
P = W(MeV/m/nC)\times q_b(nC) \times \overline{I}(=q_b/T_b)(A) \times (l_s ~(m)).
\end{equation}

Table \ref{tab1} summarizes the parameters of these structures. 
The power generated is large, on
the order of 100 MW for a 30 cm test structure with $\Delta f/f_0 ~=~ 4\%$
 and scaling quadratically
with the length of the device. 
In the 30 GHz case the dielectric device performance is
comparable to the CLIC designs \cite{R9} while being easier to manufacture and 
to incorporate parasitic mode damping.

\begin{table}
\caption{ Parameters of the 15 and 30 GHz structures and their predicted 
performance. The beam
parameters assumed correspond to those available at the CLIC test facility.}
\label{tab1} 
\begin{center}
\begin{tabular}{l l l}
\hline 
Frequency $f_0$ (GHz)&
15&
30\\
Charge per bunch $q_b$ (nC)&
10&
10\\
Bunch length $\sigma_z$ (mm)&
1 &
1\\
Bunch spacing $T_b$ (ps)&
333&
333\\
Structure length $l_s$ (cm)&
30&
30\\
Inner radius $a$ (mm)&
5& 
4\\
Outer radius $b$ (mm)&
7.38& 
5\\ 
Dielectric constant $\epsilon$&
4.5&
4.5\\
Loss tangent $\tan ~\delta$& 
0.0002&
0.0002\\
Group velocity $\beta_g$&
0.28&
0.41\\
Attenuation of the structure (dB)&
0.35& 
0.53\\
Power generated ($\Delta f/f_0 ~=~ .04$) (MW)&
94& 
178\\
Peak deceleration field $W$ (MV/m)&
23.6&
24.3\\
Normalized shunt impedance $R'/Q$ (M$\Omega$)&
0.1002&0.0516\\
\hline 
\end{tabular}
\end{center}
\end{table}

\section{Summary}

We have presented designs for dielectric power extraction structures for 
two-beam
accelerators which are able to produce power levels competitive with conventional
copper structures while at the same time possessing the advantages of ease of
construction and straightforward implementation of parasitic mode damping. We 
plan to
build and test a 15 GHz device at our laboratory and if successful build a 30 
GHz device
for the CLIC test facility.

\end{document}